\documentstyle[pre,aps,multicol,epsf]{revtex}

\begin{document}
\draft

\title{Loops in One Dimensional Random Walks}
\author{Shay Wolfling and Yacov Kantor}
\address{School of Physics and Astronomy, Tel Aviv University,
Tel Aviv 69978, Israel}
\maketitle

\begin{abstract}
Distribution of loops in a one-dimensional random walk (RW),
or, equivalently, neutral segments in a sequence of positive
and negative charges is important for understanding the low
energy states of randomly charged polymers.
We investigate numerically and analytically loops in several
types of RWs, including RWs with continuous step-length distribution.
We show that for long walks the probability density of the longest
loop becomes independent of the details of the walks and
definition of the loops.
We investigate crossovers and convergence of probability densities
to the limiting behavior, and obtain some of the analytical
properties of the universal probability density.
\end{abstract}


\begin{multicols}{2}
\narrowtext
\section{Introduction}
\label{intro}
One reason for the growing interest in polymers \cite{POL} is the desire 
to understand long chain biological macromolecules, and especially proteins. 
An important class of polymers is {\it polyampholytes} (PAs) \cite{TAN}, 
which are heteropolymers that carry a mixture of positive and negative 
charges.
In recent years, much attention has been given to the ground state
conformations of randomly charged PAs \cite{CON,KK3}.
The study of randomly charged PAs suggests that their
ground state has a structure similar to a {\it necklace}, 
made of neutral or weakly charged parts of the chain, compacting 
into globules, connected by highly charged stretched strings \cite{KK3}. 
This structure is a compromise between the tendency to reduce the 
surface area due to surface tension, and the tendency to expand due to 
the Coulomb repulsion of the total (excess) charge.
A complete analytical characterization of this structure
within the necklace model (which was obtained for homogeneously charged 
polymers \cite{DRO}), was so far not obtained. 

\begin{figure}
\centerline{\hbox{
\epsfysize=18\baselineskip
      \epsffile{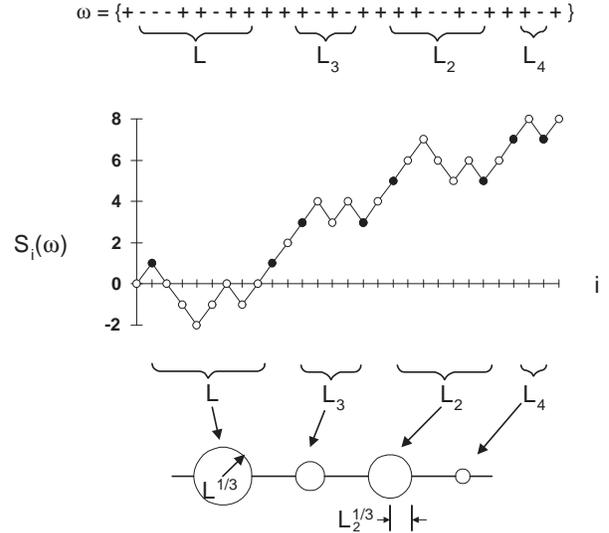}  }}
\caption [Figure.]
	 {\protect\footnotesize An example of a charge sequence $\omega$ 
	 (top), 
	 mapped into a 1-d RW $S_i(\omega)$ (middle), and a typical 
	 loops structure (bottom).
	 Filled circles indicate the starting and ending points of loops.
	 The longest loop in the RW has 8 steps ($L=8$),
	 the excess charge (which is equivalent to the total displacement
	 of the RW) is $Q=+8$, and the total length is $N=28$.}  
\label{swfig1}
\end{figure}

Since a key role in the structure of randomly charged PAs is
played by the neutral segments in the chain
(forming the beads in the necklace), Monte Carlo (MC) methods were applied 
to study their size distribution \cite{KE1,KE2,KE3,WOL}.
This problem can be investigated by mapping the charge sequence of the 
PA into a one-dimensional (1-d) RW.
The charge sequence $\omega=\{q_i\}\ (i=1,...,N;\ q_i=\pm1)$ is mapped 
into a sequence of positions $S_i(\omega)=\sum_{j=1}^i q_j$ ($S_0=0$)
of a random walker (the charges are measured in units of the
basic charge, and therefore $q_j$ is dimensionless).
The random sequence of $N$ charges is thus equivalent to an $N$-step RW,
a chain segment with an excess charge $Q$ corresponds to a
RW segment with total displacement of $Q$ steps, and
a neutral segment is equivalent to a loop inside the RW 
(see Fig.\,\ref{swfig1}).

Kantor and Erta\c{s} \cite{KE1,KE2,KE3} attempted to quantify the
necklace model, by postulating that the ground state of
a PA will consist of a single globule, formed by the longest
neutral segment of the PA, and a tail, formed by the remaining part.
However, when the longest neutral segment forms a globule, the typical
tail is very large ($\sim N$).
Since the tail itself may contain neutral segments, it will further 
fold into globules, in order to reduce the total energy \cite{WOL}.
This structure is depicted in Fig.\,\ref{swfig1}: 
The longest neutral segment contains $L$ monomers; 
it compacts into a globule of linear size proportional to $L^{1/3}$; 
in the remaining part of the chain the longest neutral segment (the 2nd 
longest neutral segment) of size $L_2$ compacts into a globule of radius
$L_2^{1/3}$, then the 3rd and so on, until the segments become very small 
(of only a few monomers). 
Eventually, all the neutral segments are exhausted and we 
are left only with strings which carry the PA's excess 
charge $Q$, and connect the globules.

There are several models in which a charge sequence is broken into 
neutral segments (similar to a RW broken into loops) \cite{MOD}.
The relation between these models and the model described above
was detailed in a previous work \cite{WOL}, but none of them
consider issues of the {\it longest} loops.

In this work we investigate the probability density $p(l)$
of a longest loop in an $N$-step RW to be of reduced length
$l\equiv L/N$, where $L$ is the number of steps in the loop.
This probability density was extensively studied 
\cite{KE1,KE2,KE3,WOL}, and some of its properties were found analytically. 
A numerical evidence for the $N$-independence of $p(l)$ in the 
$N\rightarrow\infty$ limit was presented, but a complete analytical 
description of it was not found.
We present numerical and analytical evidence that this probability
density becomes, for long walks, `universal' for a large class of RWs. 
In order to demonstrate this, we define
in section \ref{longest} the problem of the longest loop
for RWs with continuous step-length distribution, and in particular 
we define what is called a loop.
We then investigate numerically the crossover from finite $N$ to
$N\rightarrow\infty$ for different parameters defining a loop.
In section \ref{univ} we demonstrate the `universality' by
showing that the probability density
of the longest loop becomes, for $N\rightarrow\infty$,
independent of $N$, and of the parameter defining a loop, and
is the same for both RWs with steps of fixed length and steps of 
continuous length distribution.
In section \ref{prop} we investigate this `universal' probability
density, and obtain some of its properties.

\section{Longest Loops in Random Walks with Continuous Steps}
\label{longest}
In this section, unlike the models mentioned in the previous section,
we consider the problem of longest loops in
RWs with {\it continuous} step-length distribution.
According to the central limit theorem \cite{FEL}, 
the end to end distance of a RW, in which the steps are uncorrelated and
have finite variance, approaches a Gaussian probability distribution
with increasing number of steps.
In a long RW we can group a number of adjacent steps into a `rescaled step'. 
If we repeat this rescaling process, the probability distribution of a 
rescaled step length will approach a Gaussian form.
It is, therefore, convenient to start with Gaussian elementary steps,
distributed according to probability density
$f(x)=e^{-\frac{x^2}{2a^2}}/a\sqrt{2\pi}$, since the rescaling process 
does not modify the functional form of the distribution of a step
(except for increased variance).
We shall denote such a RW, as a {\it Gaussian} RW.
Although we shall address in this study Gaussian RWs,
most of the results will be more general, and 
valid for large class of RWs with continuous step-length distribution.
In Gaussian RWs, the probability density of the position $x$ of 
the `random walker' after $N$ steps is:
\begin{equation}
f(x,N)=\frac{1}{a\sqrt{2\pi N}}e^{-\frac{x^2}{2Na^2}}\ ,
\end{equation}
where $a^2$ is the variance of a single step.
Since a RW with continuous distribution of steps never
{\it exactly} returns to a previously visited position,
we say that a loop is closed if two steps in the
RW are closer than $\epsilon$ from each other.
Therefore, the probability of having a longest loop of $L$ steps in
such a RW depends on the number of steps $N$, on the standard deviation 
of the Gaussian step $a$, and on $\epsilon$ defined above. 
From dimensional arguments, it is clear that the probability of
the longest loop depends on $\epsilon$ and $a$ only through
$\epsilon/a$.

As in the problem with fixed step length, it is convenient to work with a 
probability density of the longest loop, and to explore it as a function 
of the reduced length $l$. 
The probability density is defined by 
$p(l;N,\epsilon/a)=N\cdot P(L;N,\epsilon/a)$,
where $P(L;N,\epsilon/a)$ is the probability of having a longest
loop of $L$ steps in a RW with given $a$ and $N$, when the distance 
defining a loop is $\epsilon$.

\begin{figure}
\centerline{\hbox{
\epsfysize=16\baselineskip
      \epsffile{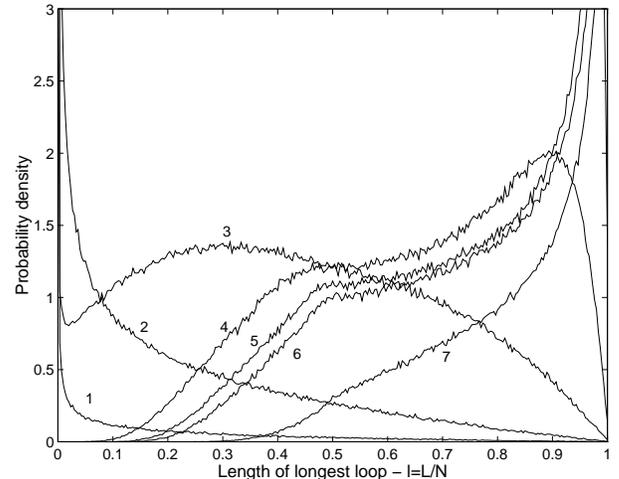}  }}
\caption [Figure.]
	 {\protect\footnotesize $p(l;N,\epsilon/a)$ {\it vs.} $l$ for 
	 several values of 
	 $\epsilon/a$:  $10^{-5}$ (1), $10^{-4}$ (2), $10^{-3}$ (3), 
	 $10^{-2}$ (4),  0.1 (5), 1 (6), 10 (7). 
	 All graphs show MC results of $10^6$ randomly selected 
	 sequences of length $N=300$.}
\label{swfig2}
\end{figure}
For a RW with steps of fixed length, it was shown numerically \cite{KE2} 
that the probability density of the longest loop
becomes $N-$independent already for modest values of $N$.
However, the probability density of the longest loop for
a RW with continuous step length distribution {\it strongly depends}
on $\epsilon$.
This dependence is depicted in Fig.\,\ref{swfig2}, showing
$p(l;N,\epsilon/a)$ for values of $\epsilon/a$ ranging from
$10^{-5}$ to 10, and for $N=300$. 
We now investigate this dependence:
The positions of steps of an $N-$step RW are spread out over a distance 
of order $a\sqrt N$.
The typical probability of a loop between two
given steps for small $\epsilon$ is of order of 
$\epsilon$ divided by the entire spread of the RW, 
i.e. $\frac{\epsilon}{a\sqrt N}$.
Since there are $\sim N^2$ pairs of steps which can form a loop,
the mean total number of loops in the $\epsilon\rightarrow0$ limit 
is of order of $N^{3/2}\epsilon/a$.
In the approximation of having no more than one loop per chain
($N^{3/2}\epsilon/a\leq1$),
it is possible to find $p(l;N,\epsilon/a)$ analytically.
Neglecting terms  ${\cal O}\left[{(\epsilon/a)}^2\right]$, 
the probability density in this single loop (s.l.) approximation is:
\begin{equation}
\begin{array}{l} 
p_{\rm s.l.}(l;N,\epsilon/a)=\\*[6pt]
\sqrt {\frac{2}{\pi}}N^{3/2}\frac{\epsilon}{a}\frac{1-l}{\sqrt l}
+\left(1-\frac{4}{3}\sqrt{\frac{2}{\pi}}N^{3/2}\frac{\epsilon}{a}\right)
\delta(l)\ .
\end{array} 
\end{equation} 
We see that when $\epsilon/a\ll 1/N^{3/2}$ no loops are formed, 
and $p(l;N,\epsilon/a)\simeq\delta(l)$.
At the opposite extreme, when $\epsilon/a\gg\sqrt{N}$, every
step closes a loop which originates at almost all the other steps in the 
walk, and therefore the first step generates a loop with the last, 
resulting in $p(l;N,\epsilon/a)\simeq\delta(l-1)$.
And indeed, in Fig.\,\ref{swfig2} we see that graph 1,
representing small $\epsilon/a$, has a strong divergence at
$l=0$, while graph 7 shows the signs of strong divergence at
$l=1$ (although $\epsilon/a$ is not very big in this case).
The intermediate values of $\epsilon/a$ exhibit crossover shapes.
In particular we note that graphs 5 and 6, where $\epsilon/a=0.1$ and
1 respectively, have a shape closely resembling the probability density 
$p(l)$ of the longest loop in RWs with steps of fixed length
\cite{KE2,WOL}, in the large $N$ limit. 

We investigated the $N-$dependence of $p(l;N,\epsilon/a)$, 
and obtained different qualitative behaviors for 
different values of $\epsilon/a$. 
For fixed $\epsilon/a<1$ the function $p(l;N,\epsilon/a)$ changes 
gradually from $\delta (l)$ towards $p(l)$ as $N$ increases. 
This behavior resembles that in Fig.\,\ref{swfig2} 
(constant $N$ and changing $\epsilon/a$), in a way that increasing $N$ 
(for constant $\epsilon/a$) is equivalent to increasing $\epsilon/a$ 
(for constant $N$). 
For fixed $\epsilon/a>1$, the function  
$p(l;N,\epsilon/a)$ changes from $\delta(l-1)$ towards $p(l)$ 
as $N$ increases, in a way that increasing $N$ (for constant $\epsilon/a$)
is equivalent to {\it decreasing} $\epsilon/a$. 
At $\epsilon/a \simeq1$, the probability density $p(l;N,\epsilon/a)$ is 
almost independent of $N$, and converges very quickly to $p(l)$.

The qualitative arguments of the previous paragraph indicate that 
for small values of $\epsilon$ the function $p(l;N,\epsilon/a)$
depends on $N$ and $\epsilon/a$ only through  $N^{3/2}\epsilon/a$.
Numerical comparison of several probability densities, having 
different values of $N$ and $\epsilon/a$ but the same value of
$N^{3/2}\epsilon/a$, confirms this dependence.
In the opposite limit $\epsilon/a\gg1$, we can find through qualitative 
rescaling arguments a single parameter for the dependence
of $p(l;N,\epsilon/a)$ on $N$ and $\epsilon/a$.
In order to represent a chain with given $N$, $\epsilon$ and $a$ as a chain
with $N^*$ `effective steps' and $\epsilon^*/a^*=1$,
we group $n={(\epsilon/a)}^2$ steps of the original chain into an 
`effective step' $a^* = a\sqrt n = a\sqrt {(\epsilon/a)^2} = \epsilon$ 
(where $\epsilon^* = \epsilon$).
Since the probability of having a loop of certain length 
(for loops longer than ${(\epsilon/a)}^2$  steps 
in the original chain) is the same in both chains, then
any chain with $N$ steps and $\epsilon/a>1$ can be divided to a chain
with $N^* = \frac{N}{{(\epsilon/a)}^2}$ steps and $\epsilon^*/a^* = 1$ 
(if $\epsilon/a$ is large enough, so that a rescaling of ${(\epsilon/a)}^2$ 
steps is meaningful, but is nevertheless smaller than $\sqrt N$, 
where $N^*$ becomes unity). 
It is therefore reasonable to assume that for
$1\ll\frac{\epsilon}{a}<\sqrt N$, the probability density of the
longest loop depends only on $N/{(\epsilon/a)^{2}}$.
Numerical comparison of several probability densities, having 
different values of $N$ and $\epsilon/a$ but the same value of
$N/{(\epsilon/a)^{2}}$, confirms this dependence.

\begin{figure}
\centerline{\hbox{
\epsfysize=16\baselineskip
      \epsffile{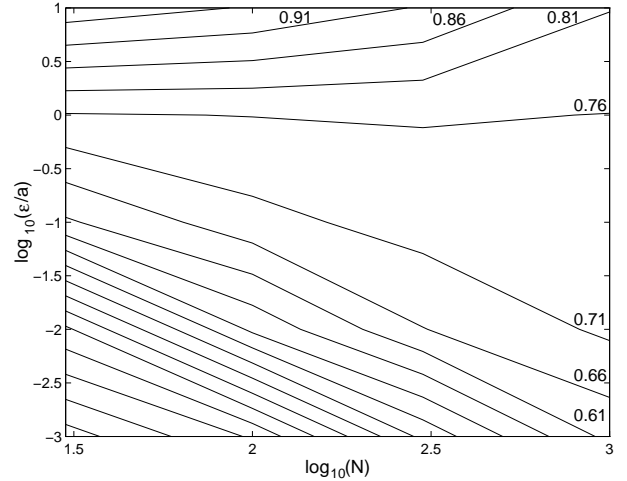}  }}
\caption [Figure.]
	 {\protect\footnotesize Lines along which $\langle l\rangle$ is 
	 constant (see text), 
	 as a function of $N$ and of $\epsilon/a$ on a logarithmic scale 
	 ($N$ = 30 to 1000 and $\epsilon/a$ = 0.001 to 10).
	 The labels above the lines are the average lengths 
	 $\langle l\rangle$ along them. The values of $\langle l\rangle$ 
	 along adjacent lines differ by 0.05.}
\label{swfig3}
\end{figure}

We want to investigate the dependence of the shapes of the different
graphs of $p(l; N,\epsilon/a)$ on $\epsilon/a$ and $N$,
and find a fixed point.
Since an infinite number of parameters is required to characterize a graph, 
we simplify the investigation by characterizing each graph by a single
parameter $\left<l\right>$ -- the average reduced length of the longest loop.
Since the functions $p(l;N,\epsilon/a)$, depicted in  Fig.\,\ref{swfig2},
are normalized functions which change gradually from $\delta(l)$, 
where $\left<l\right>=0$ to $\delta(l-1)$, where $\left< l\right>=1$, 
then $\left<l\right>$ provides a reasonable characterization of the entire 
probability density for all $l$.
Therefore, the fixed point for the entire graph of the function, is
obtained when the average $\left< l\right>$ does not change with $N$
and $\epsilon/a$.
The dependence of $\left< l\right>$ on $\epsilon/a$ and $N$ is
depicted in Fig.\,\ref{swfig3}.
The curves in Fig.\,\ref{swfig3} are lines in the `$\epsilon - N$' plane
(on a logarithmic scale), along which $\langle l\rangle$, 
(calculated by $p(l;N,\epsilon/a)$ from $10^6$ random sequences) 
has constant value. 

It is evident from Fig.\,\ref{swfig3} that the curves from both sides of 
the $\epsilon/a\simeq1$ line are not identical,
confirming that the dependence of $p(l;N,\epsilon/a)$ on $\epsilon/a$ is 
different in value between $\epsilon/a<1$ and $\epsilon/a>1$.
The value $\epsilon/a\simeq1$
is the only value of $\epsilon/a$ for which the probability density 
is (almost) the same for all $N$ (resulting in $\langle l\rangle\simeq0.76$).
Furthermore, it is evident from the figure that as $N$ increases,
$\langle l\rangle$ approaches 0.76 for all values of $\epsilon/a$.
It can be shown that as $\langle l\rangle$ approaches 0.76, the
probability density $p(l;N,\epsilon/a)$ becomes very similar to $p(l)$.
We therefore conclude, that as $N\rightarrow\infty$, the probability density 
$p(l;N,\epsilon/a)$ becomes independent of the values of $\epsilon$, $a$ 
and $N$, and very similar to $p(l)$.

In order to demonstrate that the obtained numerical results are not 
just an effect of a specific definition of what is called a loop, 
and are typical for RWs with continuous step-length distribution, 
we repeated the numerical processes for several other definitions of a loop. 
For instance, we examined the case where the distance defining a loop
is a random variable, having different values for every pair of steps.
The results obtained for these definitions were very similar to
those detailed above. 
Similar results for several definitions of a loop lead us to the notion 
that the probability  density of the longest loop is `universal', not just 
in the sense that it does not depend on the {\it values} of $\epsilon$ 
and $a$, but that it does not depend on the details of the {\it definition} 
of what is called a loop.

\section{Universality of the Probability Density of the Longest Loop}
\label{univ}
Motivated by the numerical results of the previous section, we
try to demonstrate in a more exact way, in what sense the probability 
density of the longest loop is universal.
We first show that for an $N$-step RW with steps of fixed length 
(discrete RW) this probability density becomes independent of $N$ for 
large $N$. We then generalize our results to continuous Gaussian RWs. 

\subsection{Discrete Random Walks}
\label{DRW}
In order to show that the probability density of the longest loop
for discrete RWs becomes (for long walks) independent of $N$, 
we perform a rescaling process:
We divide a given long RW with steps of fixed length into $m$ equal
sub-walks, which are `effective steps' in the rescaled walk.
Each of the $m$ sub-walks has a minimum and a maximum position of 
steps inside it.   
A loop between two sub-walks is defined as a loop in the original RW, which
starts at a step in one sub-walk, and ends at a step in the second sub-walk.
Such a loop is formed when the two sub-walks `intersect' each other
(both of them reach the same position).
This intersection occurs if either the minimum or maximum of one 
sub-walk is between the minimum and maximum of the other sub-walk.

\begin{figure}
\centerline{\hbox{
\epsfysize=16\baselineskip
      \epsffile{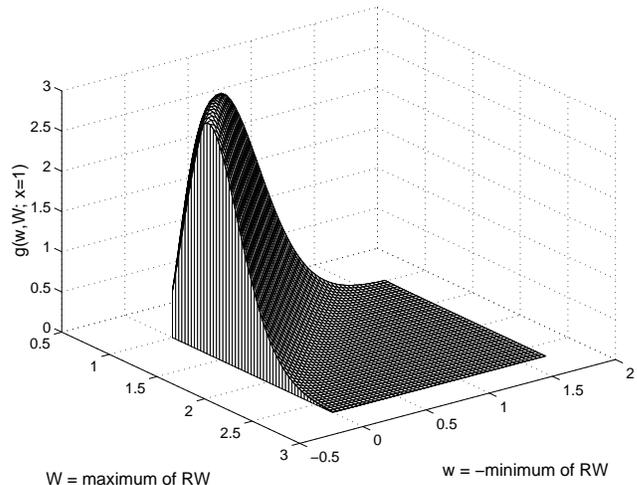}  }}
\caption [Figure.]
	 {\protect\footnotesize Mutual probability density of having a RW 
	 with a minimum $-w$ and   
	 a maximum $W$: $g(w, W; x=1)$ {\it vs.} $w$ and $W$.
	 All lengths are measured in units of the standard deviation
	 of the end position of the RW (see appendix \ref{appa}).}
\label{swfig4}
\end{figure}

The distribution of these minima and maxima depends only on the
positions of the first and last steps of each sub-walk, and therefore,
the probability of a loop between any two given sub-walks can be calculated 
analytically for any value of $N$ and $m$.
As $N\rightarrow\infty$ the functional forms of the distributions of minima 
and maxima become simple, and we therefore perform our calculation in 
that limit.
Appendix \ref{appa} describes the calculation of a
joint probability density $g(w,W;x)$ of a given RW, ending
at a position $x$ relative to its origin, to have a minimum $-w$ 
and a maximum $W$.
This probability density is valid for both discrete RWs and
continuous Gaussian RWs, and is depicted in Fig.\,\ref{swfig4}.

However, the calculation of the probability of a loop between two sub-walks 
does not enable to find the probability density of the {\it longest} loop, 
since it disregards the dependencies between the probabilities of loops. 
It is therefore convenient to turn to numerical MC methods:
We numerically generate independent RWs of $m$ `effective steps', where
the distances between their positions are distributed according to a 
Gaussian probability.
Each effective step is assigned random values of a minimum $-w$ and
a maximum $W$, according to $g(w,W;x)$, where $x$ is the distance 
between this step and the next.
We can thus obtain all the loops in the RW, and find the longest loop. 
The probability density of the longest loop is obtained by finding the 
longest loop for many independent RWs of $m$ `effective steps'.

\begin{figure}
\centerline{\hbox{
\epsfysize=18\baselineskip
      \epsffile{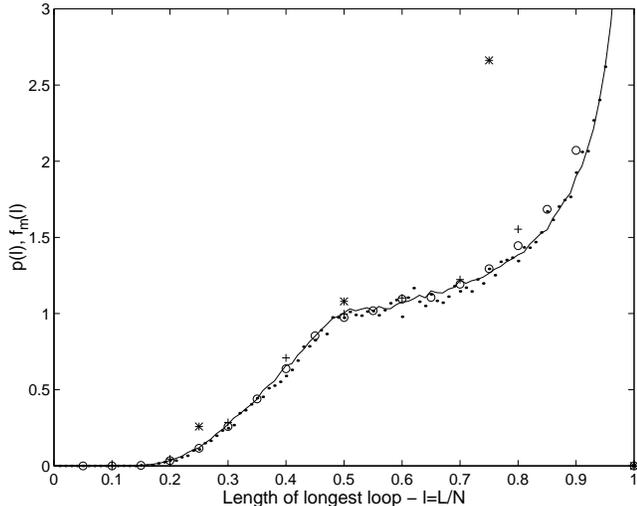}  }}
\caption [Figure.]
	 {\protect\footnotesize $f_m(l)$ {\it vs.} $l$ for several values 
	 of $m$: 4 (star), 10 
	 (plus), 20 (circle), 100 (dot) compared with $p(l)$ (line) 
	 as generated by MC simulation of $10^6$ random sequences of length 
	 $N=1000$.}
\label{swfig5}
\end{figure}

The probability density of the longest loop, emerging from the described
rescaling process, converges with increasing $m$ to the 
$N\rightarrow\infty$ limit of the probability density $p(l)$
of the longest loop for discrete RWs:
Apparently, the probability density emerging from the rescaling process
should be equal to $p(l)$, since a loop in the rescaled RW is generated if 
and only if there is a loop in the original discrete RW. 
However, given a loop between two segments in the rescaled chain, it is 
impossible to know where in the segments were the `original' steps which
generated the loop, and therefore, for each loop in the rescaled chain there
is an inherent inaccuracy of up to (plus or minus) one segment.
As $m\rightarrow\infty$ this inaccuracy vanishes.
This convergence is evident from Fig.\,\ref{swfig5}. 
We denote the probability density of having a longest loop of
length $l=L/m$, in a RW of $m$ `effective steps', by $f_m(l)$.
We have numerically obtained $f_m(l)$ for $m$=4, 10, 20, 50 and 100
(each from $10^5$ independent random sequences of length $m$), and 
compared them to $p(l)$, obtained by MC simulation of $10^6$ random
sequences of $N=1000$ discrete steps. These functions are depicted 
in Fig.\,\ref{swfig5}.
For each value of $m$, the possible values of $l$ are 
$\frac{1}{m},\frac{2}{m},\cdots,\frac{m-1}{m}$. 
The length of the longest loop is never zero, since it can be shown
that neighboring sub-walks always make a loop, and is never equal to $m$, 
since a loop from the first to the last ($m$th) step is $m-1$ segments long. 
It is evident from Fig.\,\ref{swfig5} that $f_m(l)$ converges very quickly 
with increasing $m$ to $p(l)$.

The rescaling process described in this section demonstrates
the $N-$independence (for large $N$s) of the probability density
of the longest loop, for discrete RWs: We can perform this rescaling
process to any given discrete (long) RW, and obtain the same probability
density of longest loop, independent of the number of steps $N$.
Since the rescaling process is statistically exact, the probability density 
of the longest loop, calculated from the rescaled RW, converges with 
increasing number of segments to the probability density calculated from 
the given discrete RW.
We therefore conclude that the probability density of the longest loop
for discrete RWs becomes independent of $N$ for large $N$.

\subsection{Gaussian Random Walks}
We now generalize the results, obtained in the previous sub-section
for discrete RWs, to continuous Gaussian RWs.
Since the calculation of the previous sub-section was conducted in the
$N\rightarrow\infty$ limit, the discrete nature of the RW does not affect it.
The only difference is that apparently, two segments in a rescaled 
Gaussian RW can intersect each other, where there is no
loop in the original Gaussian RW (this can happen when the positions of
steps in one sub-RW which makes a segment are distant more than $\epsilon$, 
the distance defining a `closed loop', from the positions of steps in the 
other segment).
We demonstrate that such a situation cannot occur:
We consider a Gaussian RW of $N$ steps of size $a$, and $\epsilon$ defined 
above. We scale the positions of steps of the RW by $a\sqrt N$,
making the distance between the minimum and maximum of order unity. 
The $N$ steps of the RW are spread along the position axis according to 
some probability density $q(y)$, where $\int_{\min}^{\max}q(y)dy=1$.
The average number of steps of the RW at a certain position $y$
within the (rescaled) interval defining a loop of
$\Delta y=\epsilon/(a\sqrt N)$ is given by:
\begin{equation}
N q(y) \Delta y = N q(y) \frac{\epsilon}{a\sqrt N} \sim\sqrt N\ .
\end{equation} 
The $\sqrt N$ dependence means that the average number of RW steps within 
the $\Delta y$ interval diverges with increasing $N$, for all the positions 
along the RW, and for all $\epsilon>0$.
We see that in the large $N$ limit the entire range along the position
axis is covered by the `$\epsilon-$ranges' (i.e. positions closer than 
$\epsilon$) of the steps in the Gaussian RW.
Therefore, when the minimal or maximal coordinate reached by one Gaussian 
sub-RW is between the minimum and maximum of another Gaussian sub-RW,
it is always closer than $\epsilon$ to a position of a certain step
in the second sub-RW, and a loop is formed in the original Gaussian RW.

We can therefore perform the rescaling process of the previous sub-section
on a Gaussian RW, and get a probability density of the longest loop,
which is identical to the probability density for discrete RWs.
We have thus shown that the probability density of the longest loop 
(for large enough $N$s) is independent of $N$ and is the same for both 
discrete RWs and Gaussian RWs, when a loop is defined according to an 
arbitrary $\epsilon>0$.
We therefore conclude that this probability density has some
measure of `universality'.

\section{Properties of the `Universal' Probability Density}
\label{prop}
In the previous section we numerically obtained the probability 
density of the longest loop through a rescaling process.
Although we cannot find this probability density analytically,
there are some properties and related probabilities
that can be found analytically.

\begin{figure}
\centerline{\hbox{
\epsfysize=16\baselineskip
      \epsffile{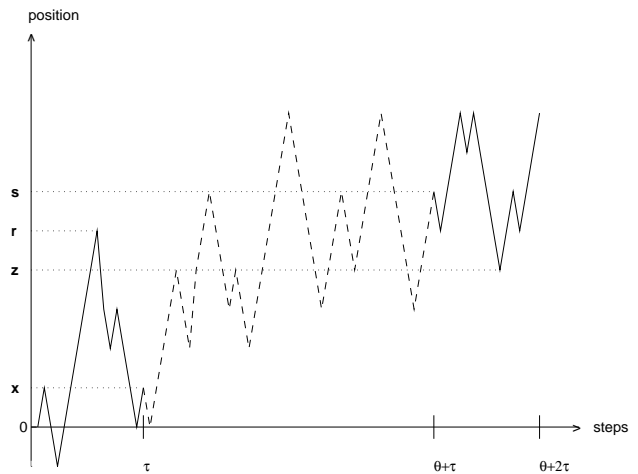}  }}
\caption [Figure.]
	 {\protect\footnotesize Two discrete RWs (solid lines), each of 
	 $\tau$ steps, 
	 which are sub-walks of a single RW (connecting dashed line).
	 The origin of the second walk is shifted $s$ along the position 
	 axis. See text for explanations of other labels.}
\label{swfig6}
\end{figure}

Since a loop between two sub-walks in a discrete 
RW is formed if the two sub-walks intersect, 
we want to calculate the probability of such intersection.
Fig.\,\ref{swfig6} depicts two such sub-RWs (solid lines), which are
part of one long RW (dashed line).
Each sub-walk is a discrete RW, having $\tau$ steps of size $a$. 
(We are interested in the limit where $\tau \rightarrow \infty$ 
and $a\rightarrow 0$ so that $a\sqrt\tau$ is finite.)
The position of the last step of the first sub-RW is $x$, while the second 
sub-RW begins at a position $s$ relative to the origin, and $\theta$ steps 
(along the original RW) after the first sub-RW ends.
The two sub-walks intersect (forming a loop in the original RW) if and 
only if $r$, the maximum of one walk (the first walk in Fig.\,\ref{swfig6})
is greater or equal to $z$, the minimum of the other walk.

These sub-walks are not independent RWs:
Fixing the position $s$ of the origin of the second sub-walk affects
the probabilities of the possible states of the first sub-walk
(and specifically the probability of the end-step position), thus
making it not completely random. This `dependence' between the
sub-walks is stronger when the number of steps $\theta$ between them
is small. In the extreme case, of neighboring sub-walks
(i.e. $\theta=0$ and the second sub-walk begins where the first one ends),
the end-position of the first sub-walk is fixed to be equal $s$.
The probability of a loop between two `dependent' sub-walks, 
separated by $\kappa$ sub-walks along the original
RW (i.e. $\theta=\kappa\tau$ in Fig.\,\ref{swfig6}),
denoted by $P_{\rm RW}(\kappa, \Delta\equiv\frac{s}{a\sqrt\tau})$, 
is derived in appendix \ref{appb} and depicted in Fig.\,\ref{swfig7}.
\begin{figure}
\centerline{\hbox{
\epsfysize=16\baselineskip
      \epsffile{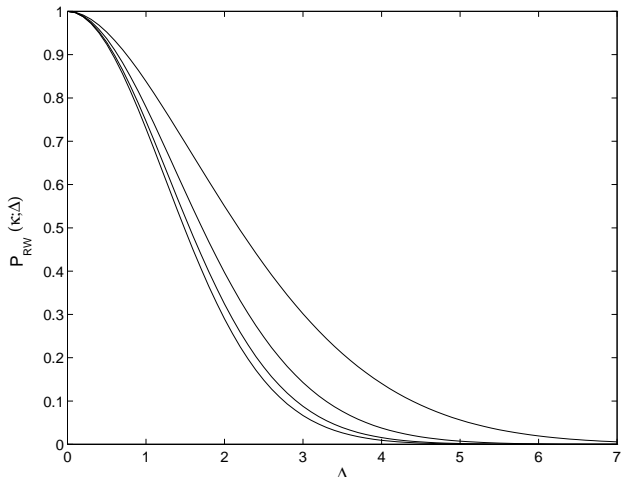}  }}
\caption [Figure]
	 {\protect\footnotesize Probability of a loop between two sub-walks 
	 $P_{\rm RW}(\kappa,\Delta)$ as a function of the normalized
	 distance between their origins $\Delta=\frac{s}{a\sqrt\tau}$
	 and the number of sub-walks between them $\kappa$ (see text).
	 Different lines for different values of $\kappa$ (from top): 
	 1, 3, 10, $\kappa\rightarrow\infty$.}
\label{swfig7}
\end{figure}

We can now calculate the probability of a loop between any two sub-walks 
in a RW, and therefore the probability of having a loop of a given length.
However, this calculation does not enable us to
find the probability density of the {\it longest} loop, since it 
disregards the dependencies between the probabilities of loops:
The fact that there is a loop (or that there are no loops) between 
two other sub-walks in the RW, changes the probability 
$P_{\rm RW}(\kappa,\Delta)$, of having a loop between two
sub-walks, resulting in an expression that we cannot find analytically.

Although we cannot find an analytical expression for $p(l)$, the probability 
density of the longest loop, we can find some of its analytical properties 
in the limit of long loops (i.e. $l\rightarrow1$).
We divide a RW with $N$ steps of fixed length $a$ into $m$ segments of 
$\tau$ steps, and investigate $f_m(\frac{m-1}{m})$, the probability of 
having a loop between the first and the last segments. 
When the last segment's origin is shifted $s$ from the origin of the RW, 
the probability of a loop between the first and last segments is
$P_{\rm RW}\left(\kappa=m-2,\ \Delta=\frac{s}{a\sqrt\tau}\right)$.
In order to obtain the probability of a loop between the first and
last segments, $P_{\rm RW}$ is integrated over all possible values of 
$s$ with the probability of the RW to reach $s$ after $m-1$ segments.
The resulting probability is:
\begin{equation}
\label{fmm1}
f_m\!\left(\!\frac{m-1}{m}\!\right)\!\!=\!\!\frac{1}{\sqrt{2\pi(m\!-\!1)}}
\int_{-\infty}^\infty \!\!\!\!\!\! d\Delta
e^{-\frac{1}{2}\frac{\Delta^2}{m-1}} 
P_{\rm RW}\left(m\!-\!2,\Delta\right)\!\!\!\ .
\end{equation} 
In the $m\rightarrow\infty$ limit, Eq.~(\ref{fmm1}) becomes:
\begin{equation}
\label{fminf}
f_m\left(\frac{m-1}{m}\right)\!=\!\frac{1}{\sqrt{2\pi m}}
\int_{-\infty}^\infty \!\!\!\!\! d\Delta
\lim_{m\rightarrow\infty}P_{\rm RW}(m,\Delta)\!\!
\simeq\!\!\frac{3.2}{\sqrt{2\pi m}}\ .
\end{equation} 
Substituting $l_0=\frac{m-1}{m}$ in Eq.~(\ref{fminf}), we get:
\begin{equation}
\label{fml0}
f_m(l_0)\simeq\frac{3.2}{\sqrt{2\pi}}\sqrt{1-l_0}\ .
\end{equation} 
The behavior of $p(l)$ in the $l\rightarrow1$ limit was obtained by
Kantor and Erta\c{s} \cite{KE2}:
\begin{equation}
\label{pul1}
p(l)=\frac{A}{\sqrt{\pi(1-l)}}\ ,
\end{equation} 
where $A$ is a numerically obtained constant ($A=1.011\pm0.001$).
This means that the probability $P_{l_0}$ of a loop to be longer than
$l_0$ is (in the $l_0\rightarrow1$ limit):
\begin{equation}
\label{Pl0}
P_{l_0}=\int_{l=l_0}^1 p(l)dl=
\frac{2}{\sqrt\pi}A\sqrt{1-l_0}\ .
\end{equation} 
It is evident (by comparing Eqs.~(\ref{fml0}) and (\ref{Pl0}) ) that the 
$l-$dependence of $f_m(l)$ is in accordance with the behavior of $p(l)$ 
in the $l\rightarrow1$ limit obtained in \cite{KE2}.

Any loop longer than $m-1$ segments in the RW begins at the first segment 
and ends at the last one, and is therefore `counted' by $f_m(\frac{m-1}{m})$.
On the other hand, all the loops that are between the first and the last 
segments of the RW are longer than $\frac{m-2}{m}$ of the chain. 
We therefore get for all $m$:
\begin{equation}
\label{ineq1}
P_{l_0=\frac{m-1}{m}} < f_m\left(\frac{m-1}{m}\right) < 
P_{l_0=\frac{m-2}{m}}\ .
\end{equation} 
In the $m\rightarrow\infty\ ,l_0\rightarrow1$ limit we get by 
substituting the definitions of $P_{l_0}$ and $f_m(\frac{m-1}{m})$ to
Eq.~(\ref{ineq1}):
\begin{equation}
\label{ineq2}
\frac{2}{\sqrt\pi}A\frac{1}{\sqrt m} < 
\frac{3.2}{\sqrt{2\pi m}} <
\frac{2}{\sqrt\pi}A\sqrt{\frac{2}{m}}\ .
\end{equation} 
From Eq.~(\ref{ineq2}) we can obtain {\it analytical} upper
and lower bounds on $A$:
\begin{equation}
\label{ineqA}
0.80<A<1.13\ ,
\end{equation} 
which are satisfied by the known numeric value of $A=1.011$ obtained in
\cite{KE2}.

\section{Conclusions and Discussion}
\label{conc}
We have defined the problem of the longest loop for RWs with continuous 
step-length distribution, investigated the resulting probability densities, 
and showed that they converge with increasing number of steps in the RW to 
the probability density of the longest loop in RWs with steps of fixed 
length.
These results motivated a rescaling process, which enabled us to obtain
a probability density of the longest loop, which is independent
of the number of steps and of the nature of the single step of the RW.
We have presented numerical and analytical evidence, suggesting that this
probability density is identical for both discrete and continuous RWs.
Investigating this `universal' probability density, we have obtained some 
of its analytical properties in the limit of long loops.
It may be possible to establish additional analytical properties of 
the problem, through further investigation of this universal function. 
However, a full renormalization--group treatment of the problem within the
derived rescaling process, in order to find a {\it complete} analytical 
solution to the problem, is expected to be quite complicated,
since we are interested in an entire probability density
and not in a single parameter.

We note that the entire rescaling process described in this work is 
applicable only to 1-d RWs. Since for RWs in dimensions $\ge2$ there is 
no analog to the minimum and maximum (there is no `boundary' to the RW), we 
cannot perform the rescaling process (for both discrete and Gaussian RWs), 
and the results do not apply to such RWs.

\begin{figure}
\centerline{\hbox{
\epsfysize=16\baselineskip
      \epsffile{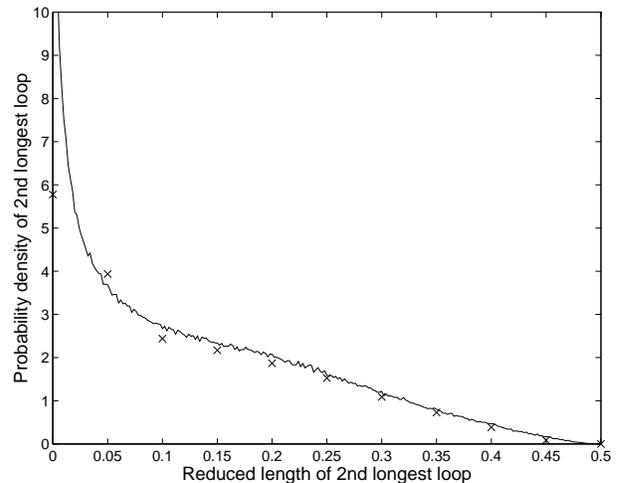}  }}
\caption [Figure.]
	 {\protect\footnotesize Probability density of the second longest 
	 loop of a rescaled chain 
	 of $m$=20 segments ($\times$), compared with the probability 
	 density of the second longest loop of a discrete RW of $N$=1000 
	 steps (line).}
\label{swfig8}
\end{figure}

The rescaling process, leading to a `universal' probability density of the 
longest loop, leads, in a similar way, to a `universal' probability density 
of the second longest loop (and third longest loop, and so on).
The definitions and processes of section \ref{univ} can be used in
order to obtain the probability density of the second longest loop:
When we performed the rescaling process, we first obtained all
the loops in the rescaled RW, and only then the longest loop.
Therefore, after `erasing' the longest loop, the second longest
loop can be obtained, and then the third, and so on.
In Fig.\,\ref{swfig8} the probability density (from $10^5$ random sequences)
of the second longest loop of a rescaled chain, having $m=20$ segments, 
is compared to the same probability density of a discrete RW of $N=1000$ 
steps, obtained by MC simulation of $10^6$ random sequences. 
As expected, these probabilities are very similar. 

The `universality' demonstrated in this study means that the probability
densities of longest loops are independent of the number of steps and of 
the nature of the single step of the RW.
Therefore, by mapping the 1-d RW into the charge sequence of a polymer,
we see that the results obtained in previous works for a specific case of 
randomly charged PAs \cite{KE1,KE2,KE3,WOL} are valid for a larger class 
of randomly charged polymers.

\section*{Acknowledgments}
This work was supported by the Israel Science Foundation under grant
No. 246/96.

\appendix
\section{Mutual Probability of a Minimum and Maximum of a Random Walk}
\label{appa}
We present the derivation of the mutual probability of
the minimum and maximum of a RW.
The probability of the position of a `random walker' satisfies the
diffusion equation, and the problem of a `random walker' bound between
a minimum and a maximum can be represented as a diffusion between
absorbing walls \cite{FEL,CHN}.
We solve the diffusion equation in one dimension \cite{KAR}:
\begin{equation} 
\frac{a^2}{2}\frac{\partial^2 f(x,t)}{\partial x^2}=
\frac{\partial f(x,t)}{\partial t}\ ,
\end{equation} 
for the probability density $f(x,t)$ of a particle starting at the origin 
and taking steps of size $a$ to be at a position $x$ after $t$ steps,
with boundary conditions of vanishing $f$ at $-w$ and $W$.
The eigenfunctions are of the form:
\begin{equation}
\sin\left(\frac{n\pi(w+x)}{w+W}\right)
e^{-\frac{1}{2}\left(\frac{n\pi}{w+W}\right)^2 a^2 t}\ ,
\end{equation}
and the prefactors are determined from the initial condition
$f(x,t=0)=\delta(x)$. 
The resulting probability density (with respect to $x$) 
$f_{w,W}(x,t)$ of a RW to be at a position $x$ after $t$ steps,
when there are absorbing walls at $-w$ and $W$, is given by:
\begin{equation}
\begin{array}{l} 
f_{w,W}(x,t)=\\*[6pt]
\frac{2}{w + W}\sum_{n=1}^{\infty}
\sin\left(\frac{n\pi w}{w+W}\right)
\sin\left(\frac{n\pi(w+x)}{w+W}\right) 
e^{-\frac{1}{2}\left(\frac{n\pi}{w+W}\right)^2 a^2 t}\ .
\end{array} 
\end{equation}
We are interested in the conditional probability of a {\it given} RW 
to have a maximum lower than $W$, and a minimum higher than $-w$. 
This conditional probability is equal to the derived probability density
$f_{w,W}(x,t)$, divided by the probability density $f(x,t)$ 
of the RW to be at a position $x$ after $t$ steps.
When measuring all distances (i.e. $w$, $W$ and $x$) in units of
$a\sqrt t$, we get for the conditional probability:
\begin{equation} 
\label{GwWx}
\begin{array}{l} 
G(w, W ;x)=\\*[6pt]
\frac{2\sqrt{2\pi}}{w+W}e^{\frac{x^2}{2}}
\sum_{n=1}^\infty
\sin\left(\frac{n\pi w}{w+W}\right)
\sin\left(\frac{n\pi(w+x)}{w+W}\right) 
e^{-\frac{1}{2}\left(\frac{n\pi}{w+W}\right)^2}\ .
\end{array} 
\end{equation} 
The derivative of $G$ in respect to $w$ and $W$:
\begin{equation} 
\label{gwWx}
g(w,W; x) =\frac{\partial^2G(w,W; x)}{\partial w\partial W} \ ,
\end{equation} 
is the mutual probability density (in respect to $w$ and $W$) of a given RW, 
ending at a position $x$, to have a minimum $-w$ and a maximum $W$.
This derivation of $g(w,W; x)$ can be repeated for a continuous
Gaussian RW with step size $a$, and therefore the mutual probability
density $g(w,W; x)$ is also valid for Gaussian RWs.
Fig.\,\ref{swfig4} in section \ref{univ} depicts this probability density.

\section{Probability of an `Intersection' Between Random Walks}
\label{appb}
We present the derivation of the probability of an `intersection'
between two RWs, that are sub-walks of one RW with steps of fixed length.
Such an intersection indicates that a loop is formed between
the two sub-walks (see Fig.\,\ref{swfig6} in section \ref{prop}).
The probability of a loop is derived for an infinitely long RW,
where each sub-walk can be treated within the Gaussian statistics.
We use the notations of section \ref{prop} -- the number of steps
in each sub-walk is $\tau$, the size of each step is $a$, there are
$\kappa\tau$ steps of the RW between the two sub-walks, the end position
of the first sub-walk is $x$, and the second sub-walk begins at a position 
$s$ relative to the origin of the first sub-walk. 
Without loss of generality we can suppose that $s\ge0$. 
The two sub-walks form a loop when the maximal coordinate of the first 
walk is greater or equal to the minimal coordinate of the second walk 
(labeled $z$).

The probability density $M(r,\tau)$ of the
maximal coordinate of a RW after $\tau$ steps can be shown 
to be (for $r>0$) twice the probability density of the
position of the RW after $\tau$ steps \cite{KAR}.
For $r<0$ the probability density $M(r,\tau)$ vanishes,
since the maximum of a RW cannot be lower than its origin.
By reflecting each RW about its origin (replacing $+a$ with $-a$
and vise versa), we see that for every RW having a maximum position of $r$, 
there is a reflected RW having a minimum position of $-r$. 
Therefore, the probability density $m(r,\tau)$ of the minimal
coordinate of a RW after $\tau$ steps equals $M(-r,\tau)$. We thus get:
\begin{equation}
\label{fi14}
M(r, \tau)=m(-r,\tau)=\left\{ 
\begin{array}{ll}
2f(r,\tau)=\frac{2e^{-\frac{r^2}{2a^2\tau}}}{a\sqrt{2\pi\tau}}
&,\ {\rm for}\ \mbox{$r\ge0$} \\
0\  &,\ {\rm for}\ \mbox{$r<0$}
\end{array}
\right.
\end{equation}
where $f(r,\tau)$ is the probability of a RW to be at a position
$r$ after $\tau$ steps.

In order to have a loop between the two sub-walks, three independent 
events must occur:
\begin{itemize}
\item[(1)] The maximal coordinate of the first sub-walk must be greater 
or equal to the minimal coordinate of the second sub-walk. The probability 
of a RW to reach a position $x$ and to have a maximum greater than $z$
after $\tau$ steps is given by \cite{FEL}:
\begin{equation} 
P_1(x,z)=
\left\{ \begin {array}{ll}
	f(2z-x, \tau) &,\ \mbox{for $z\ge\max(0,x)$}\\
	f(x,\tau) &,\ \mbox{otherwise}
	\end{array}
\right.\ \ .
\end{equation} 
\item[(2)] There is a RW between the end position of the first sub-walk
and the origin of the second sub-walk, i.e. a RW of $\kappa\tau$ steps
and a total displacement of $s-x$.
However, the position $s$ is fixed, and therefore this RW is
restricted by the existence of a RW of $\kappa\tau+\tau$ steps 
from the origin to $s$. This probability is given by
\begin{equation} 
\label{P2}
P_2(s, x)=\frac{f(s-x, \kappa\tau)}
{f(s, (\kappa+1)\tau)}\ .
\end{equation} 
\item[(3)] The minimum of the second sub-walk is equal to $z$, i.e $s-z$ 
relative to its origin. This probability is given by Eq.~(\ref{fi14}):
\begin{equation} 
P_3(s,z)\!=\!m(z\!-\!s,\tau)\!=\!\!
\left\{ 
\begin{array}{ll}
2f(s\!-\! z,\tau)\!\!\!
& \mbox{, for $z\le s$} \\
0\!\!\! & \mbox{, for $z>s$}
\end{array}
\right. \ .
\end{equation} 
\end{itemize}
The probability of a loop, denoted by $P_{\rm RW}(\kappa, \Delta)$ 
(for given $s,a,\tau$ and $\kappa$, where $\Delta\equiv s/a\sqrt\tau)$, 
is obtained by integration of $P_1\cdot P_2\cdot P_3$ over all values of 
$z$ and $x$: 
\begin{equation} 
P_{\rm RW}(\kappa, \Delta) =
\int_{z=-\infty}^{\infty}\int_{x=-\infty}^{\infty}
P_1 P_2 P_3 dz dx\ .
\end{equation} 
From the definitions of $P_1$ and $P_3$ we see that if
$z<0$ then $P_1=f(x,\tau)$. If $0\le z\le s$ then $P_1=f(x,\tau)$
when $x>z$, and if $z>s$ then $P_3$ vanishes. Therefore we get:
\begin{eqnarray}
\label{A5}
P_{\rm RW}(\kappa, \Delta) && 
=\int_{-\infty}^{0} dz P_3(s,z) \int_{-\infty}^{\infty}dx f(x,\tau)
 P_2(s,x)\nonumber\\
&&+\int_{0}^{s} dz P_3(s,z) \int_{-\infty}^{z}\!\!dx P_1(x,z) 
P_2(s,x) \\
&&+\int_{0}^{s} dz P_3(s,z) \int_{z}^{\infty}dx f(x,\tau)
P_2(s,x) \ .\nonumber\
\end{eqnarray}
The resulting probability density is depicted in Fig.\,\ref{swfig7} 
in section \ref{prop}.


\end{multicols}
\end{document}